\documentclass[lettersize,journal]{IEEEtran}
\usepackage{amsmath,amsfonts}
\usepackage{algorithmic}
\usepackage{array}
\usepackage[caption=false,font=normalsize,labelfont=sf,textfont=sf]{subfig}
\usepackage{textcomp}
\usepackage{stfloats}
\usepackage{url}
\usepackage{verbatim}
\usepackage{graphicx}
\def\BibTeX{{\rm B\kern-.05em{\sc i\kern-.025em b}\kern-.08em
    T\kern-.1667em\lower.7ex\hbox{E}\kern-.125emX}}
\usepackage{balance}
\begin{document}
\title{Remote-sensing based control of 3D magnetic fields using machine learning for in-operando applications}

\author{%
    Miguel A. Cascales Sandoval$^{1,*}$, 
    J. Jurczyk$^{1}$, 
    L. Skoric$^{2}$, 
    D. Sanz-Hern{\'a}ndez$^{3}$, 
    N. Leo$^{4,1}$, 
    A. Kovacs$^{5}$, 
    T. Schrefl$^{6,5}$, 
    A. Hierro-Rodr{\'i}guez$^{7,8,9}$, 
    and A. Fern{\'a}ndez-Pacheco$^{1,*}$\\
    \IEEEauthorblockA{$^{1}$Institute of Applied Physics, TU Wien, Wiedner Hauptstra{\ss}e 8-10, Vienna, 1040, Austria}\\
    \IEEEauthorblockA{$^{2}$University of Cambridge, Cambridge, CB3 0HE, UK}\\
    \IEEEauthorblockA{$^{3}$CNRS/Thales: Laboratoire Albert Fert, CNRS, Thales, Universit{\'e} Paris-Saclay, 1 avenue Augustin Fresnel, 91767 Palaiseau, France}\\
    \IEEEauthorblockA{$^{4}$Loughborough University, Epinal Way, Loughborough, LE11 3TU, UK}\\
    \IEEEauthorblockA{$^{5}$Department for Integrated Sensor Systems, Danube University Krems, Viktor Kaplan-Straße 2E, 2700 Wiener Neustadt, Austria}\\
    \IEEEauthorblockA{$^{6}$Christian Doppler Laboratory for Magnet Design through Physics-Informed Machine Learning, Viktor Kaplan-Straße 2E, 2700 Wiener Neustadt, Austria}\\
    \IEEEauthorblockA{$^{7}$Departamento de F{\'i}sica, Universidad de Oviedo, Oviedo, 33007, Spain}\\
    \IEEEauthorblockA{$^{8}$CINN (CSIC-Universidad de Oviedo), El Entrego, 33940, Spain}\\
    \IEEEauthorblockA{$^{9}$SUPA, School of Physics and Astronomy, University of Glasgow, Glasgow, G12 8QQ, UK}\\
    \thanks{*Corresponding author e-mails: miguel.cascales@tuwien.ac.at, amalio.fernandez-pacheco@tuwien.ac.at.}
}

\maketitle

\begin{abstract}
In-operando techniques enable real-time measurement of intricate physical properties at the micro- and nano-scale under external stimuli, allowing the study of a wide range of materials and functionalities. In nanomagnetism, in-operando techniques greatly benefit from precise three-dimensional (3D) magnetic field control, enabling access to complex magnetic states forming in systems where multiple energies are set to compete with each other. However, achieving such precision is challenging and uncommon, as specific applications impose constraints on the type and geometry of magnetic field sources, limiting their capabilities.

Here, we introduce an approach that leverages machine learning algorithms to achieve precise 3D magnetic field control using a hexapole electromagnet that is composed of three independent, non-collinear dipole electromagnets. In our experimental setup, magnetic field sensors are placed at a distance from the sample position due to inherent constraints, leading to indirect field measurements that differ from the magnetic field experienced by the sample. We find that the existing relationship between the remote and sample frames of reference is non-linear, thus requiring a more complex calibration method. To address this, we employ a multi-layer perceptron neural network that processes multiple inputs from a dynamic magnetic field sequence, effectively capturing the time-dependent non-linear field response. The network achieves high calibration accuracy and demonstrates exceptional generalization to unseen magnetic field sequences. This study highlights the significant potential of machine learning in achieving high-precision control and calibration, crucial for in-operando experiments where direct measurement at the point of interest is not possible.
\end{abstract}

\section{Introduction}

In-operando measurement techniques have become extremely valuable for exploring how nanomaterials and devices behave in real-world conditions, generally providing deeper insights compared to ex-situ methods \cite{lopez2021critical}. These techniques are employed to probe in real time the response of different materials subject to external stimuli, \textit{e.g.}, thermal catalysis assesment \cite{velisoju2023multi}, lithium-ion battery development \cite{bak2018situ}, X-ray tomography \cite{vaughan2020id15a} or growth dynamics investigation \cite{qin2024synchrotron}. In the realm of magnetism, in-operando techniques probe the live magnetic response of a system to, for instance, magnetic fields or currents. In particular, precise three-dimensional (3D) magnetic field control is essential for exploring in-depth the behavior of increasingly complex nanomagnetic devices such as the magnetic racetrack memory \cite{parkin2008magnetic}, magnetoresistive random access memory (MRAM) \cite{zhao2011design}, or magnetic sensors employed in the automotive industry for accurate position and velocity measurements \cite{hainz2019comparison}.

However, achieving such precise control is challenging, as the type of probe required for each application restricts the area around the sample, and consequently the geometry of the magnetic field source. Common probe examples include scanning tips for in-operando microscopy, electrical contacts for transport measurements, open space for optical components and light propagation, multi-axis stages, and temperature control devices. While these probes are generally compatible with dipole and quadrupole electromagnets, which can generate well-controlled variable 1D or 2D fields, extending this control to 3D while maintaining compatibility remains difficult.

\begin{figure*}[ht]
\centering
\includegraphics[scale=0.21]{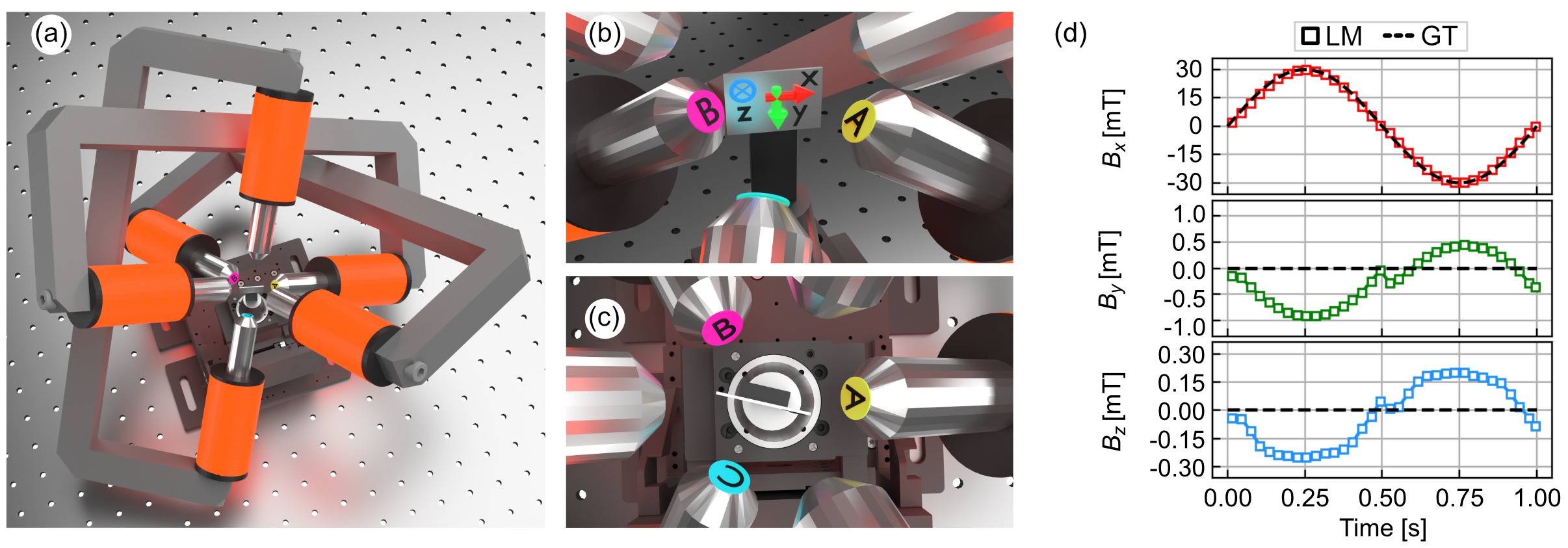}
\caption{\label{fig:figure1} (a) Top view of the hexapole design, consisting of three pairs of dipole electromagnets. (b) Zoom-in view of the sample position and FOR, represented by ($x$, $y$, $z$) coordinates. (c) Likewise for the remote FOR, composed of three independent 1D Hall-probes, denoted by (A, B, C). (d) Performance of the linear calibration using a sinusoidal signal of 30 mT amplitude along $x$, highlighting the presence of stray components in the $y$ and $z$ directions. The black-dashed line denotes the gorund truth sequence, whereas the squares denote the measured signals obtained through the linear calibration.}
\end{figure*}

An effective implementation that offers optical access and the ability to apply 3D magnetic fields is the hexapole electromagnet design described in \cite{sanz2019fabrication}, used for magneto-optical Kerr effect (MOKE) measurements. It features three pairs of dipole electromagnets arranged in a non-collinear fashion \cite{sanz2019fabrication} which enable to capture multiple reflected wave-vectors, for instance, to perform dark-field magnetometry of magnetic nanostructures \cite{sanz2023probing}. The magnetic field in this setup is measured in a "remote" frame of reference (FOR), \textit{i.e.}, separate from the "sample" FOR at the center of the magnet where the sample is placed during experiments. We find experimentally that the relationship between the fields measured in both FORs is non-linear, demanding a non-linear calibration for accurate control of the field in the sample FOR.

To model this relationship, we use a feed-forward multi-layer perceptron (MLP) neural network. The network receives a total of 9 inputs: the desired 3D magnetic field vector in the sample FOR, the 3D vector derivative representing its temporal evolution, and the previously reached maximum field 3D vector. After passing through the network's layers, it outputs the 3D vector field in the remote FOR. If the network performs correctly, reaching the field predicted in the remote FOR should yield the desired field in the local FOR. We train the network on a set of magnetic field sequences that diametrically cross our FORs center, achieving a training error of 0.08 mT per component on average. Finally, we assess the network's performance for magnetic field sequences exclusively generated for testing, finding an overall strong performance with an average testing error of 0.22 mT per component.

This paper thus presents a novel approach to perform in-operando experiments, leveraging the strengths of machine learning to overcome the difficulties arising from remote magnetic field sensing in 3D. Our results highlight the potential of neural networks to deliver the fine control over complex experimental setups, pushing forward science and technology.

\section{Results and discussion}

We use a hexapole electromagnet \cite{sanz2019fabrication}, part of a MOKE setup, for 3D magnetic field control. The hexapole, depicted in figures \ref{fig:figure1} (a,b,c), consists of three pairs of non-collinear dipole electromagnets which combined enable to apply magnetic fields along any direction of space. Figure \ref{fig:figure1} (a) provides an overview of the hexapole, with the sample positioned at the center where the axes connecting each dipole pair intersect. The optics necessary for MOKE are omitted in these figures, as they are not relevant to this discussion.

Accurate measurement and control of the vector magnetic field at the sample location and FOR $\vec{B}_{S}$ = [$B_{x}$, $B_{y}$, $B_{z}$]  (local FOR), sketched in figure \ref{fig:figure1} (b), is crucial for MOKE experiments involving 3D nanostructures \cite{sanz2023probing}. However, the design of this setup does not allow to simultaneously place a sample together with a 3-axis field sensor, \textit{i.e.}, we can not acquire MOKE signals and vector magnetic field data at the sample position at the same time. Instead, during MOKE experiments we indirectly measure the magnetic field using three independent Hall-probe sensors attached to the tip of one pole-piece from each pair, as schematically depicted in figure \ref{fig:figure1} (c). Each Hall-probe measures Hall voltage, $\vec{V}_{R}$ = [$V_{A}$, $V_{B}$, $V_{C}$] (remote FOR), directly proportional to the axial component of the local magnetic field at each pole piece.

\begin{figure*}[ht]
\centering
\includegraphics[scale=0.22]{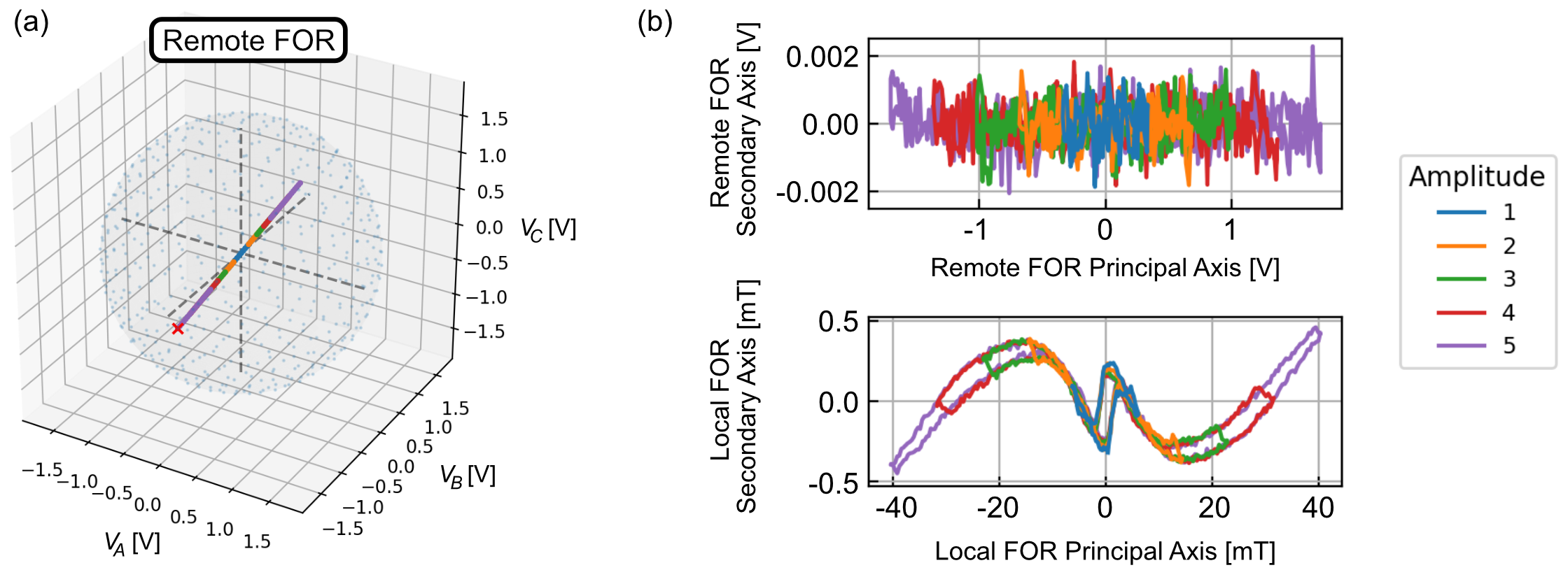}
\caption{\label{fig:figure2} (a) 3D representation of the directions sampled for training using Fibonacci's algorithm. The radial nature of the sequences is exemplified for just one direction, marked by the red "X". The amplitude colormap is shared with panel (b). (b) Comparison between the signals measured in both remote and local frames of reference for the red "X" direction of figure (a), plotted on their principal component axes (directions of highest variance) for direct comparison. The figure emphasizes the non-linear and hysteretic behavior observed in the local frame of reference, which explains why the linear matrix calibration approach lacks precision.}
\end{figure*}

To calibrate the remote FOR to measure in the local FOR, we need to find the relationship between the two. To achieve this, we place a 3-axis magnetic field sensor aligned with the ($x$, $y$, $z$) axes of the sample FOR, giving us $\vec{B}_{S}$. We then simultaneously collect data for various magnetic fields, obtaining pairs of vectors in the two FORs. As Maxwell's equations are linear in free space, both frames of reference would be in principle related through a linear transformation, \textit{i.e.}, via a $3\times 3$ linear matrix (\textbf{M}) that accounts for both rotation and scaling, satisfying $\vec{V}_{R} = \textbf{M}\cdot \vec{B}_{S}$. We experimentally obtain the matrix's coefficients via a linear regression fit between the datasets (refer to supplementary material for details). In practice, this is the first out of a two-step process; we first use the matrix to convert the field of interest from the local to the remote FOR, and then we optimize the currents outputted to the electromagnet to obtain the desired field in the remote FOR. For this, we have developed a proportional-integral-derivative (PID) feedback algorithm (not shown here) that deals with the magnet's non-linear and hysteretical behavior achieving the desired field.

Figure \ref{fig:figure1} (d) illustrates the performance of the linear matrix calibration for a typical sinusoidal signal of nominal 30 mT amplitude along $x$, with zero amplitude along $y$ and $z$. Measurement of the real field in $\vec{B}_{S}$ via the 3-axis sensor reveals that the linear calibration lacks precision, as significant unwanted stray components arise deviating from the nominal ground truth signal. For instance, the $y$ component exhibits a peak stray component of the order of 1 mT, which is approximately 3$\%$ of the intended amplitude in $x$. Stray components could lead to strange unexpected magnetic behaviors or unexpected switches, making very difficult to interpret MOKE hysteresis loops.

These results thus show a non-linear relationship between the remote and sample FORs. This may be expected in this case, caused by the hysteresis of the electromagnet’s core magnetic  material, combined with cross-talk between the pole pieces. Here, cross-talk refers to the influence of one dipole electromagnet on the magnetic state of the others, resulting from the large separation between pole-piece tips ($\approx$ 3.5 cm) relative to their diameter ($\approx$ 0.8 cm), consequently reducing the degree of field channeling and uniformity within the air gap. Although closer and thicker pole pieces would produce more homogeneous fields and reduce cross-talk, the system is designed this way to allow a clear optical path and sufficient working space around the sample area.

To capture the non-linear relationship between FORs we utilize a MLP neural network, which offers a powerful approach to model the existing intricate relationships. We collect our training data in such a way we cover as uniformly as possible the entire 3D space, as shown in figure \ref{fig:figure2}, using Fibonacci's algorithm \cite{gonzalez2010measurement} to generate evenly spaced points on a spherical surface. Each of these points denotes a direction in space, on which we perform a magnetic field sequence which diametrically crosses through the center. At each of these directions we perform several sequences at different amplitudes, as exemplified in figure \ref{fig:figure2} (a) for the direction marked by the red "X" symbol (in the remainder of the manuscript, we refer to these sequences as radial). To ensure that we obtain the correct fields, we utilize our PID feedback algorithm acting on the remote FOR.

In figure \ref{fig:figure2}, we show examples of radial signals simultaneously measured by both remote and local FORs, plotted on their principal component axes \cite{kurita2019principal} for direct comparison. The principal component analysis is done respectively and independently for the time varying sequences of both FORs, [$V_{A}$, $V_{B}$, $V_{C}$] (t) and  [$B_{x}$, $B_{y}$,  $B_{z}$] (t). Strikingly, what appears as a near-perfect line in the remote FOR (considering noise levels) does not translate to a perfect line in the local FOR. Instead, there is a hysteretic, non-linear behavior, underscoring why a linear matrix approach fails to capture this dependence, and highlighting further the value of using a neural network. Although not shown here, similar behavior occurs when the feedback control is used to optimize the field in the local FOR: line-shaped signals emerge in the local FOR, while hysteresis appears in the remote FOR. To thoroughly probe both FORs, we perform the data acquisition twice; once optimizing the remote FOR and measuring hysteresis in the local, and again optimizing the local FOR while measuring hysteresis in remote.

We probe in total 527 directions in space to achieve the most uniform sampling of the 3D sphere, measuring five triangular sequences at each direction with a fixed field sweep rate of dH/dt = 140 mT/s and amplitudes of [8, 16, 24, 32, 40] mT. Before training, we apply 25-point binning to our dataset to reduce noise levels and thereby enhance the noise robustness of our network. This binning, combined with the maximum 10 kHz data acquisition rate of our digital-to-analog converter, results in an effective sampling rate of 400 Hz, yielding approximately 1.4 million samples, each consisting of two vectors that describe the same magnetic field in the two FORs.

\begin{figure*}[ht]
\centering
\includegraphics[scale=0.24]{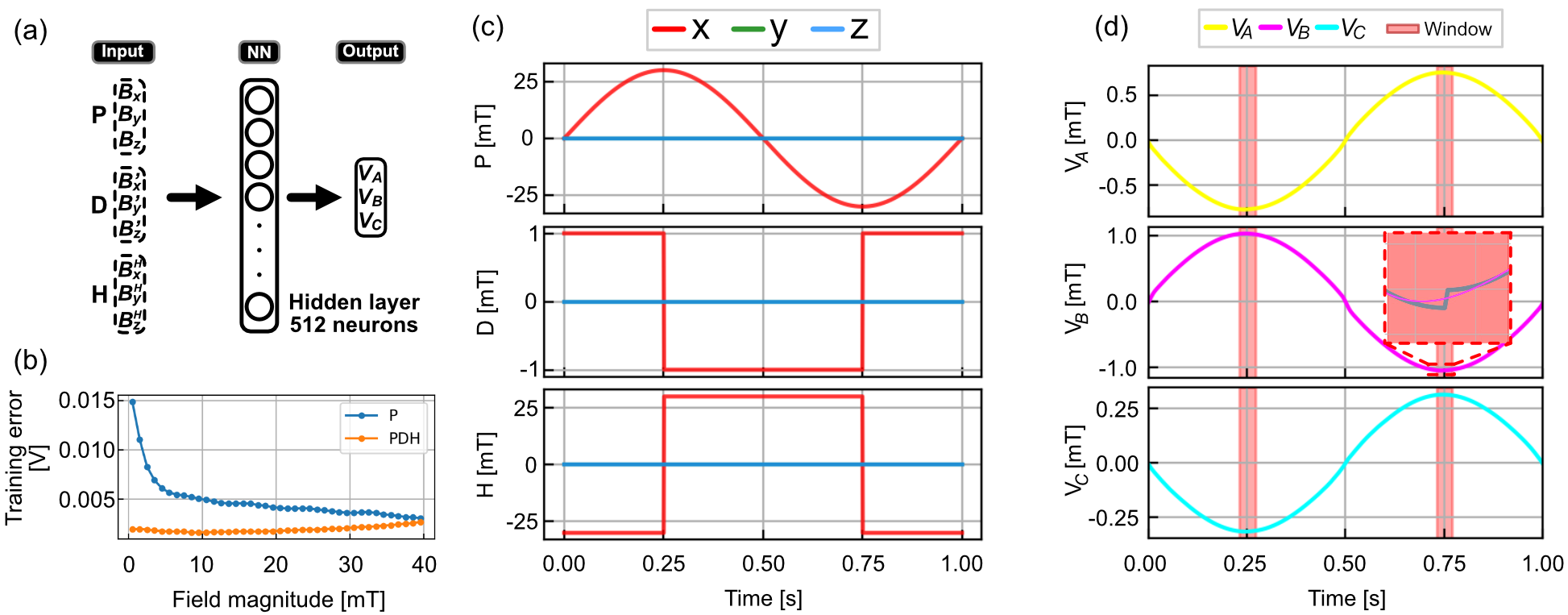}
\caption{\label{fig:figure3} (a) Illustration of the MLP neural network architecture's inputs, hidden layer, and output. "P" represents the point of interest in the sample FOR, whereas "D" an "H" enclose the variational and historical time-dependence. The blue and orange colors represent the inputs given to each of the two networks discussed here. (b) Plot of mean-absolute error as a function of field magnitude, comparing the trained networks P and PDH. (c) Example inputs to the trained PDH network, showing a sinusoidal signal in the "P" panel, the normalized backward difference in the "D" panel, and the magnetic history in the "H" panel. The ($x$, $y$, $z$) colormap is shared amongst the three panels. (d) Plot per component of the PDH network output in the remote FOR. The discontinuity is highlighted in panel $V_{B}$ by plotting the original points in slate-gray color. Window refers to the points removed, and later bridged through the cubic spline.}
\end{figure*}

We use a supervised learning approach, in which the network takes the 3D magnetic field in the local FOR $\vec{B}_{S}$ as input, and outputs the corresponding magnetic field in the remote FOR $\vec{V}_{R}$. We then apply PID control to optimize the network’s output in the remote FOR, which if it works as expected, will yield the desired field in the local FOR, matching the network's input field. We thus first train a network that takes $\vec{B}_{S}$ as input and outputs $\vec{V}_{R}$, labeled as "P" in figure \ref{fig:figure3} (a). After experimenting with various combinations of neurons and hidden layers (without systematic hyperparameter tuning), we select a structure with a single hidden layer of 512 neurons and ReLU activation. We use a learning rate of 0.001 with the Adam optimizer from the Keras library \cite{chollet2015keras}, the mean-squared error metric for the loss, and train for 150 epochs. We show in figure \ref{fig:figure3}(b) the mean-absolute error after training as a function of the field magnitude in the remote FOR, revealing a clear trend: the training error decreases as field magnitude increases. Interestingly, the error rises sharply near zero field magnitude, indicating this as a particular point. We use this metric instead of the mean-squared error used for the actual training in this graph, in order to visualize the error in volt units.

The sharp rise at low fields is understood by analyzing figure \ref{fig:figure2} (b): in the remote frame of reference, the magnetic field path follows roughly the same points as it moves forward and backward, essentially forming a line. However, the corresponding field in the local frame does not retrace the same path; instead, it exhibits hysteresis. This results in a degenerate relationship between the two frames of reference, where at the same field point in the remote frame, there are two possible field values in the local frame, depending on the direction of the field sweep. Consequently, the network finds a solution that averages between the two hysteresis branches, which is generally larger at field magnitudes close to zero.

To address the issue of the time-dependent hysteretic response, we would need to incorporate temporal information into the network's input. This is not a straightforward task, as the sequences generated evolve over time, but the network evaluates sequences point by point. Thus, to capture the temporal dynamics, we compute the backward difference $\vec{B}^{'}_{S}$ at the point of interest $\vec{B}_{S}$ and give this additional vector as input, helping the network distinguish between the different branches of the hysteresis. To ensure generality, we normalize this vector to eliminate any dependence on the field sweep rate. We refer to this as "D" in figure \ref{fig:figure3} (a). Furthermore, as shown in figure \ref{fig:figure2} (b), varying field amplitudes in the remote frame of reference lead to different closure behaviors in the local frame. To assist the network further in learning how to consistently remove the hysteresis, we also provide the maximum field previously reached per component $\vec{B}^{H}_{S}$. This input is called "H" in figure \ref{fig:figure3} (a). We refer to this network with 9 inputs as PDH.

We train PDH using the same hyperparameters as for P, showing the corresponding mean-absolute errors for both trained networks as a function of field magnitude in figure \ref{fig:figure3} (b). Including these additional parameters as input reduces training errors across the entire field magnitude range, with significant improvement near zero field magnitude, where hysteresis is generally larger. We thus select the PDH network for the remainder of this manuscript.

We next illustrate an example of the inputs and outputs of the trained PDH network in figures \ref{fig:figure3} (c,d). We use the waveform labeled as P in figure \ref{fig:figure3} (c); a sinusoidal signal with a 30 mT amplitude in the $x$ direction, with the other two components set to zero. The "D" panel displays the normalized backward difference of the signal from panel "P". Finally, panel "H" shows the magnetic history vector, completing the 9 inputs that the PDH network utilizes. The signals are discretized at a rate of 1 kHz.

Figure \ref{fig:figure3} (d) shows the outputted three components in the remote frame of reference. Notably, where "D" changes sign, the output exhibits a small jump between consecutive points in some cases. We have explored various machine learning approaches to address this discontinuity, such as adding regularization or penalties to minimize differences between closely spaced points, avoiding normalization of the backward difference, and employing recurrent neural networks (RNNs), specifically using long short-term memory (LSTM) cells. However, this undesired feature did not significantly change. To remove this potentially problematic high-frequency component, we develope an algorithm that fits a cubic spline after removing the points near the discontinuity, ensuring a smooth curve.

In the final section, we analyze the network's performance on sequences generated exclusively for testing. We generate them by creating waveforms in the local FOR (discussed below), calculating their P, D, and H values, and then using the trained PDH network to predict the field in the remote FOR. We then use PID control to optimize the predicted fields in the remote FOR, while monitoring the error between the measured field in the sample FOR and the input sequence provided to the network.

\begin{figure*}[ht]
\centering
\includegraphics[scale=0.19]{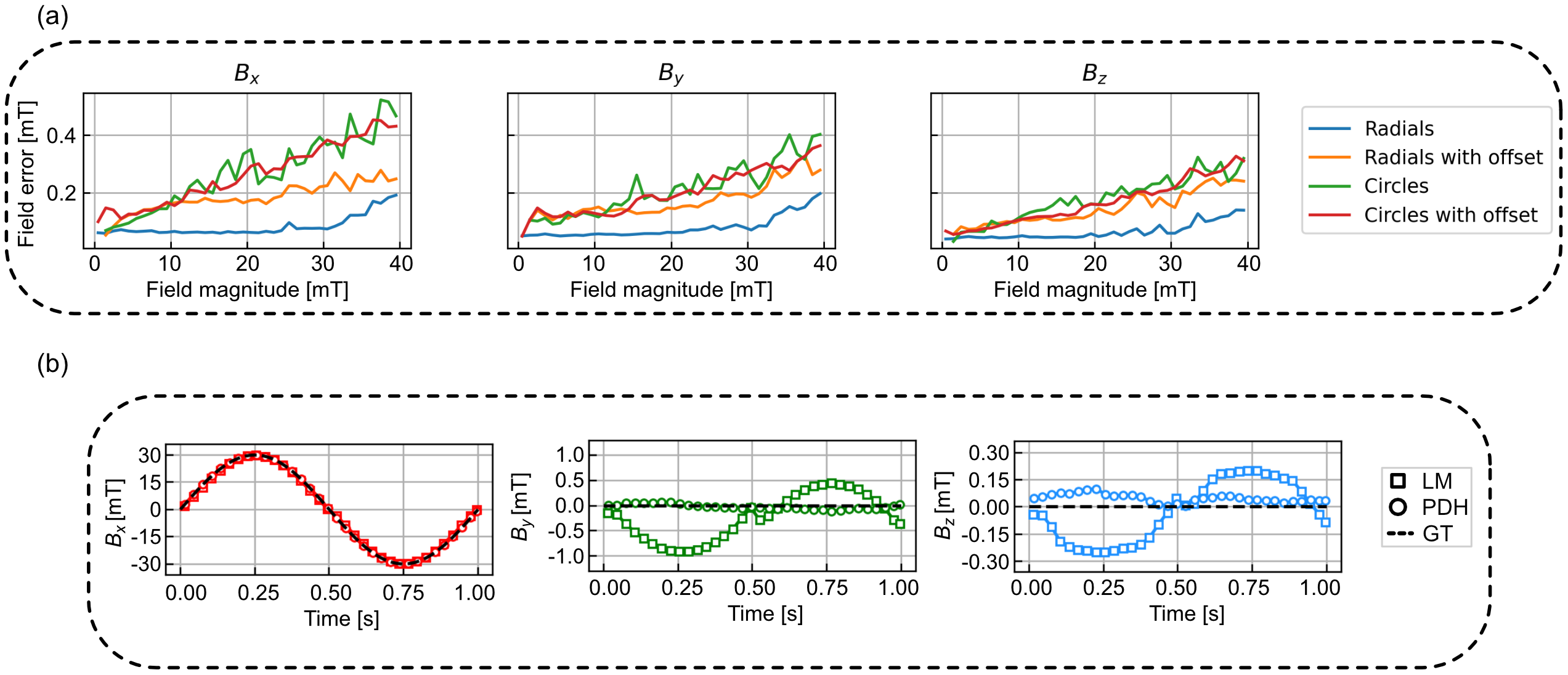}
\caption{\label{fig:figure4} a) Evaluation of the PDH network's performance using test data from various sequences, showing the network's ability to generalize and accurately predict magnetic field components across different field configurations. (b) Comparison between the linear matrix calibration and the PDH network's output.}
\end{figure*}

The first type of waveforms in the testing dataset consist of radial sequences similar to those used in training, but oriented in 50 random directions different from the ones used in training. In each of these directions, we apply four random amplitudes combined with various field sweep rates. Next, we create another set of radial sequences, to which we add random constant field offsets in different directions. Finally, we generate circular sequences with varying amplitudes on different planes, some centered on (0,0,0), while others are displaced by random offsets. These are the most distinct from the ones used in training, challenging the network’s generalization since their "D" component is always orthogonal to that of the radial training sequences.

The errors for the different testing sequences are plotted in figure \ref{fig:figure4} (a) as a function of the field magnitude, for each field component and for each sequence type. The first conclusion is that the sequences with the lowest errors are the radials, as expected, as even though they are not along the training directions, they are the most similar. The error is almost flat up to 30 mT, on average under 0.08 mT per component. The fact that the error starts to rise roughly above 30 mT is due to the fact that the density of points at those amplitudes reduces drastically. However, even in that case it remains under 0.2 mT throughout the whole range.

The test sequences with the next lowest error are the radials with random offsets. This is also expected, as these sequences are more similar to the training datasets than the circular ones. The average error associated to these sequences is 0.18 mT, with a peak of 0.3 mT at the end of the field range. The testing errors for both circular sequences are very similar regardless of the offsets, which is explained by the fact that these sequences are completely different from the radials, thus the offsets become a second order effect. The quality of the results remains remarkable even for these sequences; the network manages to extrapolate with great precision, maintaining errors per component under 0.22 mT on average, with the peak near 0.4 mT at the end of the range. Even this extreme case already improves significantly the \textbf{M} calibration.

To conclude the testing analysis, we show in figure \ref{fig:figure4} (b) the signals obtained both with the linear matrix (same as in figure \ref{fig:figure1}) and with the trained network PDH, clearly demonstrating how the neural network is able to reduce the unwanted components by approximately one order of magnitude.

\section{Conclusions}

In this work we detail how machine learning, specifically a MLP neural network, enables precise calibration of a hexapole electromagnet for precise control of 3D magnetic fields in an in-operando experimental setup. The network effectively deals with non-linearities inherent to the remote-sensing scheme, relating measurement and sample FORs. These non-linearities arise from cross-talk between poles and the magnetic core's hysteretic, non-linear nature.

Our findings indicate that incorporating temporal information into the network's input, specifically adding normalized differences "D" and maximum field history "H" alongside the desired magnetic field "P", significantly improves the network's ability to handle such hysteresis and non-linear effects. This approach reduces errors across the full field magnitude range, particularly near zero field, where the simpler linear matrix method is less effective. Further, testing demonstrates the network's generalization capability: the PDH network consistently maintains errors below 0.08 mT per component for radial field sequences, and below 0.22 mT for circular ones. These results highlight the potential of neural networks in in-operando nanodevice experiments, in cases where a limited access to the sample results in nonlinear remote control over the measurements.

\section{Acknowledgments}

This work was supported by the European Community under the Horizon 2020 Program, Contract No. 101001290 (3DNANOMAG). The raw data supporting the findings of this study will be openly available at a repository from TU Wien. A.H.-R. acknowledges support by the Spanish MICIN (Refs. MICIN/AEI/10.13039/501100011033/FEDER, UE under grants PID2019-104604RB and PID2022-136784NB) and by Asturias FICYT (grant AYUD/2021/51185) with the support of FEDER funds. L.S. acknowledges support from the EPSRC Cambridge NanoDTC EP/L015978/1. For the purpose of open access, the author(s) has applied a Creative Commons Attribution (CC-BY) licence to any Author Accepted Manuscript version arising from this submission. The financial support by the Austrian Federal Ministry of Labour and Economy, the National Foundation for Research, Technology and Development and the Christian Doppler Research Association is gratefully acknowledged.

\bibliographystyle{IEEEtran}
\bibliography{bibliography}

\end{document}